\documentstyle[PASJadd,epsf]{PASJ95}
\draft

\newcommand{\beq}{\begin{equation}}
\newcommand{\eeq}{\end{equation}}
\newcommand{\beqa}{\begin{eqnarray}}
\newcommand{\eeqa}{\end{eqnarray}}

\newcommand{\vol}{}
\newcommand{\no}{}
\newcommand{\lett}{}
\newcommand{\spage}{1}
\newcommand{\rdate}{}
\newcommand{\adate}{}

\begin{document}

\title{
\begin{flushright}
\normalsize
OU-TAP-59 \\
submitted to PASJ \\
March 1997
\end{flushright}
\ \\
Fokker-Planck Models of Star Clusters\\
with Anisotropic Velocity Distributions\\
III. Multi-Mass Clusters
}
\author{
Koji {\sc Takahashi}\\
{\it Department of Earth and Space Science,
Graduate School of Science, Osaka University,}\\
{\it Toyonaka, Osaka 560}\\
{\it E-mail: takahasi@vega.ess.sci.osaka-u.ac.jp}}

\abst{
The evolution of globular clusters
driven by two-body relaxation is investigated 
by means of numerical integration of the two-dimensional
Fokker-Planck equation in energy--angular momentum space.
The two-dimensional Fokker-Planck equation allows the development of velocity anisotropy.
We include a spectrum of stellar masses in this paper.
The radial anisotropy develops, that is, the radial velocity dispersion exceeds the tangential one,
in the outer halo of multi-mass clusters as in single-mass clusters.
However, the evolution of the velocity anisotropy depends significantly
on the stellar mass in some cases.
In fact the tangential velocity dispersion becomes dominant
around the half-mass radius for massive components 
in clusters with a steep mass function.
The development of this tangential anisotropy is closely related
to the initial cooling of the massive components toward energy equipartition.
Our simulation results indicate that multi-mass anisotropic King-Michie models
are not always appropriate 
for describing the velocity anisotropy in globular clusters.
}

\kword{Clusters: globular ---  Fokker-Planck equation ---
 Numerical methods --- Stars: stellar dynamics --- Velocity anisotropy}

\maketitle
\thispagestyle{headings}


\section{Introduction}\label{sec:intro}

This is the third in a series 
of studies on the dynamical evolution of globular star clusters 
by using Fokker-Planck (FP) models which allow velocity anisotropy.
In the preceding two papers, 
pre-collapse (Takahashi 1995, Paper I) and post-collapse evolution (Takahashi 1996, Paper II) of isolated, single-mass clusters was studied.
In this paper, as a natural extension of those works, 
we study the evolution of multi-mass clusters 
(clusters with a stellar-mass function).
We especially pay attention to the development of the velocity anisotropy of each mass-component.

Direct numerical integration of the FP equation (Cohn 1979, 1980)
has been the most important tool of the study of globular cluster evolution.
Various factors such as the mass function, the heating effects of binaries and the galactic tidal field,
which are expected to be important for the evolution of real globular clusters,
were incorporated into FP simulations.
However, velocity anisotropy was usually neglected
because of numerical difficulties in the integration of the two-dimensional FP equation (Cohn 1985; Paper I).
Most of the difficulties can be removed by using the improved integration scheme described in Paper I\@.

For the above reason,
although the evolution of multi-mass clusters has been investigated by many authors
(e.g., Inagaki, Saslaw 1985; Chernoff, Weinberg 1990; Murphy et al.\ 1990;
Lee et al.\ 1991)
by using {\it isotropic} FP models,
the development of the anisotropy in multi-mass clusters has seldom been discussed.
Only recently, it was discussed by Spurzem and Takahashi (1995)
as well as Giersz and Heggie (1996) to some extent.
Spurzem and Takahashi (1995) compared anisotropic gaseous models,
isotropic FP models, and $N$-body models for the evolution of
two-component clusters.
At that time anisotropic FP models were not available.
Giersz and Heggie (1996) performed $N$-body simulations for small-$N$
multi-mass clusters ($N$=250, 500, and 1000).
All of their simulations started with the same power-law initial mass
function.
They combined statistically the results from many simulations.
However, it is still difficult to see the evolution of the anisotropy 
because of large statistical fluctuations, 
especially for massive components comprising very small numbers of
particles.
In this paper we present the results of anisotropic FP simulations for simple
two-component clusters as well as clusters with a realistic power-law
mass function.
FP simulations do not suffer from the problem of statistical
fluctuations.
Our results clearly show how the evolution of the anisotropy depends on the
stellar mass and the initial mass function.

In fitting observational data to theoretical models,
King models (King 1966) and their extensions have been used most frequently.
The mass function and velocity anisotropy are incorporated in
multi-mass King-Michie models (e.g., Gunn, Griffin 1979).
In these models
the distribution function of each component depends on the specific angular momentum $J$ through the factor $\exp [-J^2/(2r_{\rm a}^2\sigma^2)]$, 
where $r_a$ is the anisotropy radius and $\sigma$ is the velocity dispersion;
radial anisotropy is significant for $r \gtsim r_{\rm a}$.
The choice of this form of angular momentum dependence is rather arbitrary.
For example, $r_{\rm a}$ can depend on mass components,
but it is usually set identical for all components for model simplicity.
Gaseous model simulations by Spurzem and Takahashi (1995) and
our FP simulations show that the states of velocity anisotropies are significantly different between different components in some cases.
In fact the tangential velocity dispersion becomes dominant
around the half-mass radius for massive components in a cluster with a steep mass function.
Therefore, anisotropic King-Michie models are not always
useful extensions of isotropic King models.

In section 2, the basic equations and initial conditions are described.
The results of our simulations are presented in section 3.
The conclusions and discussion are given in section 4.
In the Appendix
the diffusion coefficients for the multi-mass anisotropic FP equation are given.

\section{Fokker-Planck Models}

\subsection{Basic Equations}

We assume that star clusters are spherically symmetric and isolated from
other systems.
We take account of the mass function of stars.
For computational convenience, the mass function is represented by 
a set of discrete mass components.
The number density of the $i$th component 
(of the individual stellar mass $m_i$) in $\mu$-space is denoted by 
$f_i({\bf r},{\bf v},t)$.
In a spherical system in dynamical equilibrium, the distribution function is a function of only 
the energy per unit mass ($E$) and the modulus of the angular momentum per unit mass ($J$).
We define the scaled angular momentum $R$ as
\beq
R = \frac{J^2}{J_{\rm c}^2(E)} \,,
\eeq
where $J_{\rm c}(E)$ is the angular momentum of a circular orbit of energy $E$.
The number density of the $i$th component in $(E,R)$-space, $N_i(E,R)$,
 is given by
\beqa
  N_i(E,R) &=& 4\pi^2 P(E,R) J_{\rm c}^2(E) f_i(E,R) \nonumber \\
         &\equiv& A(E,R)f_i(E,R)
\eeqa
(Cohn 1979).
Here, $P(E,R)$ is the orbital period,
\beq
P(E,R) \equiv 
2\int_{r_{\rm p}}^{r_{\rm a}} \frac{dr}{v_{\rm r}} \,,
\eeq
where $v_{\rm r}=\{2[\phi(r)-E]-J^2/r^2\}^{1/2}$,
$\phi(r)$ is the gravitational potential,
and $r_{\rm p}$ and $r_{\rm a}$ are the pericenter and apocenter radii, respectively.

The multi-mass FP equation under a fixed 
potential can be written in a flux-conserving form (cf. Cohn 1979),
\beq
A \frac{\partial f_i}{\partial t} =
- \frac{\partial F_{E i}}{\partial E} - \frac{\partial F_{R i}}{\partial R}
\label{eq:fp} \,,
\eeq
where
\beqa
-F_{E i} & = &   D_{E i} f_i
+ D_{EE i} \frac{\partial f_i}{\partial E} 
+ D_{ER i}\frac{\partial f_i}{\partial R} 
\,, \nonumber \\
-F_{R i} & = &   D_{R i} f_i 
+ D_{RE i}\frac{\partial f_i}{\partial E} 
+ D_{RR i}\frac{\partial f_i}{\partial R} \label{eq:flux} \,.
\eeqa
The expressions for coefficients $D_{EE i}, D_{E i}$, etc. are given in the Appendix.

The gravitational potential is determined from Poisson's equation,
\beqa
\nabla^2\phi 
&=& - 4\pi G \sum_i \rho_i    \nonumber \\
&=& - 4\pi G \sum_i m_i \int f_i d^3{\bf v} \,,
\eeqa
where $G$ is the gravitational constant.

We include the heating effects by three-body binaries
and use the heating rate formula of Lee et al. (1991,
see also Drukier et al.\ 1992; Grabhorn 1992).
In their formulation, 
the total heating rate per unit volume by three-body binaries is given by
\beq
\dot{E}_{\rm tot} 
= C_{\rm b} G^5 \left( \sum_i \frac{n_i m_i^2}{\sigma_i^3} \right)^3 
\sigma_0^2 \,,
\eeq
where $n_i$, $m_i$, and $\sigma_i$ are the number density, stellar mass, and one-dimensional velocity dispersion, respectively, of the $i$th component,
and $\sigma_0^2$ is the density-weighted central velocity dispersion.
The constant $C_{\rm b}$ was set to 90 in this study as usual.
The total heating rate is distributed to each component in proportion to the mass density such that
\beq
\dot{E}_i = \frac{\rho_i}{\rho_{\rm tot}} \dot{E}_{\rm tot} \,,
\eeq
where $\rho_{\rm tot}$ is the total density.
The orbit-averaged heating rate (per unit mass) 
for each component used in the FP equation is given by
\beq
\langle \dot{E}_i \rangle_{\rm orb} 
= \left. \int_{r_{\rm p}}^{r_{\rm a}} \frac{dr}{v_{\rm r}} 
\frac{\dot{E}_{\rm tot}}{\rho_{\rm tot}}
\right/ \int_{r_{\rm p}}^{r_{\rm a}} \frac{dr}{v_{\rm r}} \,. \label{eq:oahr}
\eeq

\subsection{Multi-Mass Models}

We chose multi-mass Plummer's model (e.g. Spitzer 1987, p13) as the initial cluster model.
The distribution function of the $i$th component is
\begin{equation}
f_i(E,R)=\frac{24\sqrt{2}}{7\pi^3}\frac{r_0^2}{G^5M^5}\frac{M_i}{m_i}E^{7/2},
\end{equation}
where $r_0$ is a scale radius, $M$ the total mass of the cluster, and $M_i$ the
total mass of the $i$th component.
In the model, the velocity distribution is isotropic everywhere,
and the velocity dispersions for all components are equal, 
i.e., there is no mass-segregation.
In real globular clusters, 
just after the dynamical equilibrium was established after their
formation, 
the velocity distribution was probably more or less anisotropic.
If globular clusters formed through the cold collapse of proto-clusters,
the radial velocity dispersion might exceed the tangential one on the
average.
We consider, however, only initially isotropic models in this study for
simplicity.
The equipartition of the velocity dispersions seems reasonable 
initial conditions,
since violent relaxation does not produce mass segregation (Lynden-Bell 1967).


First, we consider two-component models.
They are simple, but very useful to understand the essential
features of the evolution of multi-component clusters.
Parameters of the models we studied are listed in table 1.
Each model is specified by three parameters, $m_2/m_1$, $M_2/M_1$, 
and $N$ (total number of stars).
The evolution of similar models was calculated 
by using anisotropic gaseous models by Spurzem and Takahashi (1995).

Next, we consider continuous mass-function models.
Let ${\cal N}(m)\,dm$ be the number of stars in the mass interval
$(m,m+dm)$.
We chose a simple power-law mass function,
\beq
{\cal N}(m)\,dm \propto m^{-\alpha} \,dm, \qquad
m_{\rm min} \leq m \leq m_{\rm max} \label{eq:mf} \,.
\eeq
We use $K$ discrete mass components,
and give the stellar mass of the $i$th component by
\beq
m_i = m_{\rm min}\left(\frac{m_{\rm max}}{m_{\rm min}}
      \right)^{(i-\frac{1}{2})/K} \qquad (i=1, 2, \dots, K) \,.
\eeq
The total mass of the $i$th component is
\beq
M_i=\int_{m_{i-\frac{1}{2}}}^{m_{i+\frac{1}{2}}} {\cal N}(m)m\,dm \,.
\eeq
Parameters of the power-law mass-function models are listed in table 2.
In all the models $m_{\rm max}/m_{\rm min}=10$ and the total number of
stars $N=10^5$.
Only the power-law index of the mass function, $\alpha$, is different
among the three models:
$\alpha=$1.5, 2.5, and 3.5 for models C1, C2, and C3
(Salpeter's mass function is $\alpha=2.35$).
We used $K=$10 components in all the simulations in the present study.
This component number is not very large,
but it is sufficient, at least, to see the qualitative features of the
evolution of clusters with the continuous mass function (cf. Chernoff,
Weinberg 1990).

\section{Results}

The results are presented in units such that
$G=M=1$ and ${\cal E}_{\rm i}=1/4$, where 
${\cal E}_{\rm i}$ is the initial total binding energy
of the cluster.
The mean half-mass relaxation time $t_{\rm rh}$ is defined by
(Spitzer 1987, p40)
\beq
t_{\rm rh} = 0.138\frac{N^{1/2}r_{\rm h}^{3/2}}{\bar{m}^{1/2}G^{1/2} \ln
\Lambda},
\eeq
where $N$ is the total number of stars, 
$r_{\rm h}$ is the radius containing half of the total mass,
$\bar{m}=M/N$ is the mean stellar mass,
and $\ln \Lambda$ is the Coulomb logarithm.
We set $\Lambda=0.11N$ (Giersz, Heggie 1994).
Time is expressed in units of the initial half-mass relaxation time $t_{\rm rh,i}$ for each model.

\subsection{Two-Component Clusters}


Figure 1a shows the evolution of the Lagrangian radii containing
1, 2, 5, 10, 20, 30, 40, 50, 75, and 90\% of the total mass of each component
for model T1.
Figures 1b and c are the same as figure 1a, but for models T2 and T3, respectively.
It is clear that mass segregation proceeds during the pre-collapse stage in each model.
The mass segregation develops most prominently 
and the core collapse occurs most rapidly 
(in units of the initial half-mass relaxation time $
t_{\rm rh,i}$ for each model)
in model T3.
During the post-collapse stage, on the other hand, 
the mass segregation almost stops to proceed and the cluster expands nearly self-similarly (Murphy et al.\ 1990; Giersz, Heggie 1996).


Figures 2a1 and b1 show the evolution of the temperature, $T_i \equiv m_i \sigma_i^2$, where $\sigma_i$ is the {\it mean} one-dimensional velocity dispersion, 
at the Lagrangian radii containing 1, 10, 20, 50, and 90\% of the total mass of the cluster ($M$), for model T1.
Figures 2a2 and b2 are for model T2, and figures 2a3 and b3 are for model T3.
The temperature of the heavy component at the central region rapidly decreases at first in every model.
This is due to the approach to the equipartition of energy by two-body relaxation.
The degree of the temperature decrease is largest in model T3,
and next largest in model T2.
The heavy component can lose the energy most easily in model T3, 
because there are enough amount of light stars which receive the energy 
and because the initial temperature difference between the two components is largest.


We measure the degree of velocity anisotropy by a quantity
\begin{equation}
\beta_i \equiv 1 - \sigma_{{\rm t}i}^2 / \sigma_{{\rm r}i}^2 \,,
\end{equation}
where $\sigma_{{\rm r}i}$ and $\sigma_{{\rm t}i}$ is the radial and tangential one-dimensional velocity dispersions, respectively, of the $i$th component.
Note that this anisotropy parameter is different from 
the anisotropy parameter $A$ used in Papers I and II by a factor of 2.
Figures 3a1 and b1 show the evolution of $\beta_1$ and $\beta_2$, respectively, 
 at the Lagrangian radii of the total mass of the cluster for model T1.
Figures 3a2 and b2 are for model T2, and figures 3a3 and 3b3 are for model T3.

In every model, the anisotropy of the light component develops in a similar way as in single-mass clusters (cf. figure 3 of Paper II).
The anisotropy does not penetrate into so inner regions in model T1 as in models T2 and T3,
because the light component does not collapse very much in model T1
as shown in figure 1a.
The development of the anisotropy of the heavy component is more interesting, 
especially, in models T2 and T3.
$\beta_2$ at the inner regions becomes negative 
(i.e. {\it tangential anisotropy} develops) at the early stages
(cf. Spurzem, Takahashi 1995).
This phenomenon is most clearly seen in model T3;
$\beta_2$ at the half-mass radius attains the minimum value of $-0.5$,
and $\beta_2$ becomes significantly negative even at the 90\%-mass radius.
It may be a surprise that such large tangential anisotropy appears at the outer part of the cluster.
The tangential anisotropy of the heavy component appears also in model T1 temporarily,
though its degree is very small.
As Spurzem and Takahashi (1995) discussed and we explain right below
again,
the appearance of the tangential anisotropy is closely related to the initial temperature differences between the mass components 
and subsequent cooling of the heavy component.


Figure 4 shows the profiles of the radial and tangential temperatures,
$T_{{\rm r}i}$ and $T_{{\rm t}i}$,
at $t=1.5t_{\rm rh,i}$ in model T3.
Here we define the radial and tangential temperatures as
$T_{{\rm r}i} \equiv m_i \sigma_{{\rm r}i}^2$ and
$T_{{\rm t}i} \equiv m_i \sigma_{{\rm t}i}^2$.
For the heavy component, there are big humps in the profiles, 
i.e., temperature inversions of large amplitudes appear.
Recall that the velocity dispersions of the two components are equal at the initial time.
At the early stage, the temperature of the heavy component drops to attain the equipartition of the energy, as seen in figure 2.
Since the local relaxation time increases with radius due to the density decrease,
the temperature drops most rapidly at the center.
Therefore, it can happen that the temperature decrease at regions around the core cannot catch up with the core temperature decrease,
and then a temperature inversion appears.
[Temperature inversions can appear also during post-collapse evolution 
and trigger gravothermal oscillations (Bettwieser, Sugimoto 1984).]
Comparing the radial and tangential temperatures, 
a larger temperature inversion appears in the tangential temperature,
because circular-orbit stars do not go further into the central region 
and hardly reflect the central temperature decrease.
Therefore, the tangential velocity dispersion exceeds the radial one at this temperature-inversion region.
This situation is clearly seen in figure 4.
As the degree of the initial temperature decrease is larger,
the tangential anisotropy develops more.
Thus most significant tangential anisotropy develops in model T3.


Though we have already shown the evolution of the anisotropy at the Lagrangian radii in figure 3,
it is also instructive to show the evolution of the radial profile of the anisotropy.
This is shown in figures 5 and 6 for models T1 and T3.
The core radius $r_{\rm c}$ and the half-mass radius $r_{\rm h}$ are also indicated in the figures for reference.
Here we define the core radius as
\begin{equation}
r_{\rm c} \equiv \sqrt{\frac{9\sigma_0^2}{4\pi G\rho_0}} \,,
\end{equation}
where $\rho_0$ is the total central density
and $\sigma_0$ is the density-weighted central velocity dispersion
(cf. Spitzer 1987, p16).
In model T1, the anisotropy profiles of the two components are always similar.
On the other hand, in model T3, they are very different.
The anisotropy profile of the light component in model T3 is simple and similar to that in model T1.
As we already stated, 
the most striking feature in the anisotropy of the heavy component 
is the initial development of the tangential anisotropy outside the core.
The subsequent relaxation process produces high-energy heavy stars on radial orbits
and tends to erase the tangential anisotropy.
However, its trace can be observed 
as a hollow in the anisotropy profile
around the half-mass radius 
even after several tens of the initial half-mass relaxation time.


We also computed isotropic models corresponding to anisotropic models T1, T2 and T3 for comparison.
Mass segregation proceeds essentially in a similar way as in the anisotropic models.
Although rather strong anisotropy develops in the anisotropic models as shown in figure 3,
the evolution of the {\it mean} velocity dispersions (temperatures) in the anisotropic models is not very different from that in the isotropic models!
Figure 7 shows the evolution of the temperature of each component
at the Lagrangian radii containing 1, 10, 20, 50, and 90\% of the total mass of the cluster, for the isotropic model corresponding to model T3.
One noticeable difference between the isotropic and anisotropic models 
appears in the temperature at the 90\%-mass radius;
it is lower in the anisotropic model.
This is because the halo is more expanded in anisotropic models (Papers I
and II).

\subsection{Clusters with A Continuous Mass Function}


Figures 8a, b, and c show the evolution of the central density of each component in models C1, C2, and C3.
We note that gravothermal oscillations occurred in model C1
when we used small enough time step which was comparable to the central
relaxation time (cf. Cohn et al.\ 1989; Breeden et al.\ 1994).
However, figure 8a presents the result of the computation in which longer time step was
used so as to suppress gravothermal oscillations artificially.
The use of the long time step largely reduced computational time and
enabled us to follow long-term evolution easily.
The evolution of the outer parts of the cluster is not affected by gravothermal oscillations very much,
and neither is the mean trend of the long-term core evolution
(cf. Lee et al.\ 1991).
In every model, mass segregation proceeds in the pre-collapse phase
and the central region soon becomes dominated by the massive components.
In the post-collapse expansion phase, however,
mass segregation almost stops to proceed 
as in the cases of two-component clusters.


Figures 9, 10, and 11 show the evolution of the anisotropy profile of each component in models C1, C2, and C3.
Qualitative characteristics of the anisotropy evolution are the same as 
those in two-component models.
The development of the anisotropies of light components is simple;
radial anisotropies ($\beta>0$) develop and $\beta$ increases with the radius except for the outermost regions.
The development of the anisotropies of heavy components is more complicated;
these radial profiles fluctuate,
and strong tangential anisotropies ($\beta<0$) develop in models with a
steep mass-function such as C2 and C3.
The development of tangential anisotropies is directly related to 
the initial temperature decrease toward energy equipartition, 
as we described in section 3.1.
The degree of the initial temperature decrease is largest in model C3, 
because the relative amount of heavy components is smallest.
The anisotropy $\beta$ of the heaviest component attains the minimum value of 
$\beta \approx -1$ ($\sigma_{\rm t}^2/\sigma_{\rm r}^2 \approx 2$) in model C3
around the half-mass radius.
We also note that perfect isotropy is always established inside the core radius in all the models.

In model C1 the difference in the anisotropy profile between mass components is relatively small.
In model C3, on the other hand, the anisotropy profile strongly depends on the stellar mass.
The radial anisotropy develops for light components and the tangential anisotropy does for heavy components.
As a natural consequence of this,
for intermediate-mass components
the development of the anisotropy is more or less suppressed.


Figures 12a and b show the profiles of the radial and tangential temperatures
of each component 
at $t=2.4t_{\rm rh,i}$ in model C3.
The radial temperature profiles are almost normal;
only slight temperature inversions appear in the heaviest two components.
The tangential temperature profiles of heavy components are rather abnormal;
there is a region where the tangential temperature of the heaviest component increases outward by a factor of about two.

\section{Conclusions and Discussion}

We have studied the development of velocity anisotropy in multi-mass star clusters by solving the FP equation in energy--angular momentum space.
Simulations were performed for simple two-component clusters 
as well as clusters with a power-law mass function (ten components were employed).
The initial cluster models were multi-mass Plummer's models,
where the velocity distribution is isotropic
and the velocity dispersions for all components are equal.
The equipartition of the velocity dispersions is a probable initial condition for globular clusters,
and implies that the temperature is inversely proportional to the stellar mass.
Therefore, the temperatures of heavy components decrease rapidly at the early evolutionary stage, 
and this leads to the development of the tangential anisotropy,
that is, the tangential velocity dispersion exceeds the radial velocity dispersion.
The tangential anisotropy appears more prominently in clusters with a steeper mass function,
because the degree of the initial temperature decrease is larger for such clusters.

In model C3, a cluster with a moderately steep mass-function ($\alpha=3.5$),
the maximum tangential anisotropy such as $\sigma_{\rm t}^2/\sigma_{\rm r}^2 \approx 2$ for the heaviest component appears around the half-mass radius.
Such strong tangential anisotropy is seen for a long time
of the order of $10t_{\rm rh,i}$.
This means that it may have survived until the present time,
if we adopt the typical half-mass relaxation time for the {\it present} Galactic globular clusters, $10^9$ yr, for $t_{\rm rh,i}$.
If the strong tangential anisotropy has survived, 
can we obtain any observational evidence of it?
To the author's knowledge, no positive evidence has been reported,
although the direct measurement of the velocity anisotropy is very difficult.
Unfortunately,
the most massive stars are non-luminous degenerate remnants,
i.e. neutron stars or heavy white dwarfs, in the present-day clusters.
(Of course they were once more massive.)
Fortunately, on the other hand, the next heaviest stars are 
bright red giants.
They may show significant tangential anisotropy in some globular clusters.
In any case we should at least note that anisotropy profiles may be more complex 
than simple theoretical models, such as King-Michie models, predict.
As we mentioned in section 1,
multi-mass King-Michie models are most popular anisotropic models used in model-fitting of observational data.
They cannot reproduce anisotropy profiles such as those observed in models C2 and C3.
(However, this does not necessarily means that the density profiles cannot be well fitted to King-Michie models.)

We have not considered the effects of stellar evolution in this study.
Primordial high-mass stars have already lost all or almost all of their
initial masses.
The mass loss has significant effects on the early evolution of 
globular clusters;
weakly bound clusters are destroyed (Chernoff, Weinberg 1990).
The way in which the anisotropy of each component develops depends on
the stellar mass and the mass function as we described above.
Therefore it may be also affected by stellar evolution.
To a first approximation, however, we may assume that stellar evolution
has the effect of changing only the initial condition,
because the main-sequence time scale of primordial massive stars are much 
shorter than the time scale of relaxation (Lee et al.\ 1991).

In this study we have considered globular clusters to be isolated systems.
The effects of tidal truncation of clusters by the Galactic potential 
was investigated for single-mass clusters 
in a separate paper (Takahashi et al.\ 1997).
The paper showed that 
the radial anisotropy in the halo is highly depressed during the
post-collapse evolution due to rapid loss of radial-orbit stars. 
We will study more realistic models for the dynamical evolution 
of globular clusters in the Galaxy
including both the tidal truncation and the mass function.

\par
\vspace{1pc}\par
The author thanks Professor Hyung Mok Lee for useful comments.
This work was supported in part by the Grant-in-Aid for Encouragement of Young
Scientists by the Ministry of Education, Science, Sports and Culture of Japan 
(No. 1338).

\setcounter{equation}{0}
\renewcommand{\theequation}{A\arabic{equation}}

\section*{Appendix.\ The Diffusion Coefficients}

The expressions for the diffusion coefficients appearing in the single-mass FP equation were given by Cohn (1979).
The extension to multi-mass cases is easy.
In this appendix we give the expressions for the coefficients in
the multi-mass FP equation for completeness.
We use basically the same notations as those used by Cohn (1979).

The coefficients in equation (\ref{eq:flux}) are given by
\beqa
D_{E i} &=&  -8\pi^2 J_{\rm c}^2 \sum_j m_i m_j 
\int_{r_{\rm p}}^{r_{\rm a}} \frac{dr}{v_{\rm r}} F_{1 j} \,, \nonumber \\
D_{R i} &=& -16\pi^2 R r_{\rm c}^2 \sum_j m_i m_j 
\int_{r_{\rm p}}^{r_{\rm a}} \frac{dr}{v_{\rm r}} 
\left( 1-\frac{v_{\rm c}^2}{v^2} \right) F_{1 j}\,, \nonumber \\
D_{EE i} &=& \frac{8\pi^2}{3} J_{\rm c}^2 \sum_j m_j^2 
\int_{r_{\rm p}}^{r_{\rm a}} \frac{dr}{v_{\rm r}} v^2 (F_{0 j}+F_{2 j}) \,, \nonumber \\
D_{ER i} &=& D_{RE i} = \frac{16\pi^2}{3} J^2 \sum_j m_j^2
\int_{r_{\rm p}}^{r_{\rm a}} \frac{dr}{v_{\rm r}}
 \left( \frac{v^2}{v_{\rm c}^2}-1 \right) (F_{0 j}+F_{2 j}) \,, \nonumber \\
D_{RR i} &=& \frac{16\pi^2}{3} R \sum_j m_j^2 
\int_{r_{\rm p}}^{r_{\rm a}} \frac{dr}{v_{\rm r}}
\Bigg\{ 2\frac{r^2}{v^2} \left[ v_{\rm t}^2 
\left( \frac{v^2}{v_{\rm c}^2}-1 \right)^2 +v_{\rm r}^2 \right] F_{0 j} 
\nonumber \\
&& +3\frac{r^2v_{\rm r}^2}{v^2} F_{1 j}
+\frac{r^2}{v^2} \left[ 2v_{\rm t}^2 
\left( \frac{v^2}{v_{\rm c}^2}-1 \right)^2 -v_{\rm r}^2 \right] F_{2 j}
\Bigg\} \,,
\eeqa
where $v_{\rm c}(E)$ is the speed of a circular-orbit star of energy $E$,
$v^2=2[\phi(r)-E]$, $v_{\rm t}^2=J^2/r^2$, and $v_{\rm r}^2=v^2-v_{\rm t}^2$.
The functions $F_{0 j}$, $F_{1 j}$, and $F_{2 j}$ are defined by
\beqa
F_{0 j} (E,r) &=& 4\pi\gamma \int_0^E dE'\, \bar{f}_j(E',r) \,, \nonumber \\
F_{1 j} (E,r) &=& 4\pi\gamma \int_E^\phi dE'\, \bar{f}_j(E',r) 
\left( \frac{\phi-E'}{\phi-E} \right)^{1/2} \,, \nonumber \\
F_{2 j} (E,r) &=& 4\pi\gamma \int_E^\phi dE'\, \bar{f}_j(E',r) 
\left( \frac{\phi-E'}{\phi-E} \right)^{3/2} \,,
\eeqa
where $\gamma = 4\pi G^2 \ln \Lambda$.
The function $\bar{f}_j(E',r)$ is an assumed isotropic background distribution,
 and defined by
\beq
\bar{f}_j(E,r) = \frac{1}{2R_{\rm max}^{1/2}} \int_0^{R_{\rm max}}
\frac{dR}{(R_{\rm max}-R)^{1/2}} f_j(E,R)  \,,
\eeq
where $R_{\rm max}(E,r) = 2r^2[\phi(r)-E]/J_{\rm c}^2(E)$ is the maximum
allowed value of $R$ for all the orbits of energy $E$ which pass through
radius $r$.


\section*{References}

\re
Bettwieser E., Sugimoto D.\ 1984, MNRAS 208, 493
\re
Breeden J.L., Cohn H.N., Hut P.\ 1994, ApJ 421, 195
\re 
Chernoff D.F., Weinberg M.D.\ 1990, ApJ 351, 121
\re 
Cohn H.\ 1979, ApJ 234, 1036
\re 
Cohn H.\ 1980, ApJ 242, 765
\re 
Cohn H.\ 1985, in Dynamics of Star Clusters, IAU Symp No.113, ed
J.~Goodman, P.~Hut (D.~Reidel Publishing Company, Dordrecht) p161
\re 
Cohn H., Hut P., Wise M.\ 1989, ApJ 342, 814
\re 
Drukier G.A., Fahlman G.G., Richer H.B.\ 1992, ApJ 386, 106
\re
Giersz M., Heggie D.C.\ 1994, MNRAS 268, 257
\re
Giersz M., Heggie D.C.\ 1996, MNRAS 279, 1037
\re
Grabhorn R.P., Cohn H.N., Lugger P.M., Murphy B.W.\ 1992, ApJ 392, 86
\re
Gunn J.E., Griffin R.F.\ 1979, AJ 84, 752
\re
King I.R.\ 1966, AJ 71, 64
\re
Lee H.M., Fahlman G.G., Richer H.B.\ 1991, ApJ 366, 455
\re
Lynden-Bell D.\ 1967, MNRAS 136, 101
\re
Murphy B.W., Cohn H.N., Hut P.\ 1990, MNRAS 245, 335
\re 
Spitzer L.\ Jr 1987, Dynamical Evolution of Globular Clusters (Princeton
University Press, Princeton)
\re 
Spurzem R., Takahashi K.\ 1995, MNRAS 272, 772
\re 
Takahashi K.\ 1995, PASJ 47, 561 (Paper I)
\re 
Takahashi K.\ 1996, PASJ 48, 691 (Paper II)
\re
Takahashi K., Lee H.M., Inagaki S.\ 1997, submitted to MNRAS

\label{last}

\clearpage

\begin{table}[htb]
\begin{center}
Table~1.\hspace{4pt}Models of two-component clusters.\\
\vspace{6pt}
\begin{tabular}{cccc} \hline
Model       &  $m_2/m_1$ & $M_2/M_1$ & N    \\ \hline
T1          &  2       & 1           & 5000 \\
T2          &  2       & 1/9         & 5000 \\
T3          &  5       & 1/9         & 5000 \\ \hline
\end{tabular}
\end{center}
\end{table}

\bigskip

\begin{table}[htb]
\begin{center}
Table~2.\hspace{4pt}Models of power-law mass-function clusters.\\
\vspace{6pt}
\begin{tabular}{cccc} \hline
Model       & $m_{\rm max}/m_{\rm min}$ & $\alpha$ & N        \\ \hline
C1          & 10        & 1.5      & $10^5$    \\
C2          & 10        & 2.5      & $10^5$    \\
C3          & 10        & 3.5      & $10^5$    \\ \hline
\end{tabular}
\end{center}
\end{table}

\newpage


\begin{center}
\leavevmode
\epsfverbosetrue
\epsfxsize=14cm \epsfbox{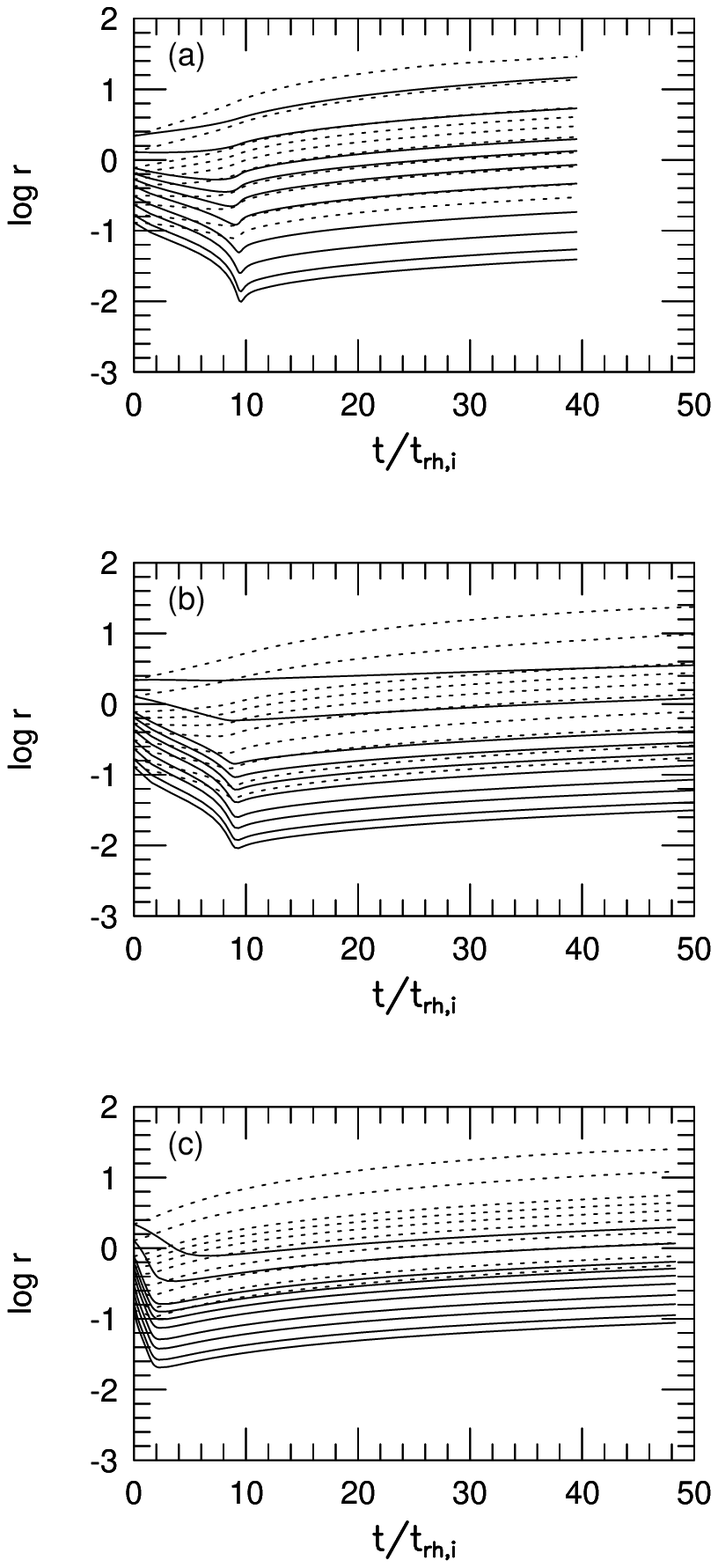}
\end{center}
\begin{Fv}{1}
{30pc}
{
Evolution of the Lagrangian radii containing
1, 2, 5, 10, 20, 30, 40, 50, 75, and 90\% of the total mass of each component for models (a) T1, (b) T2, and (c) T3.
The solid lines represent the heavy component, 
and the dotted lines represent the light component.
}
\end{Fv}

\newpage

\begin{center}
\leavevmode
\epsfverbosetrue
\epsfxsize=14cm \epsfbox{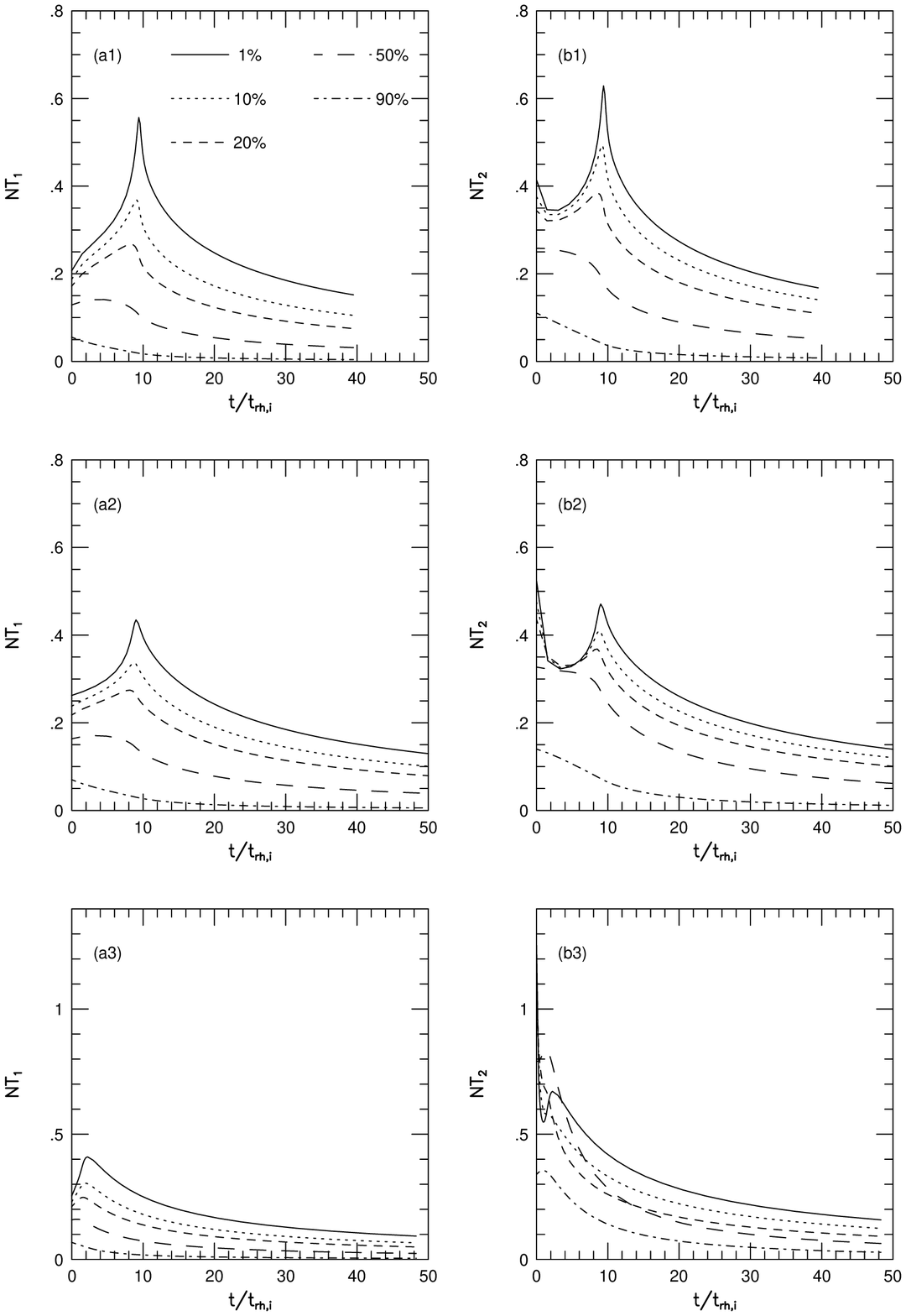}
\end{center}
\begin{Fv}{2}
{30pc}
{
(a1) Evolution of the temperature of the light component $T_1$ 
at the Lagrangian radii containing 1, 10, 20, 50, and 90\% of the total mass of the cluster, for model T1.
The ordinate is the temperature multiplied by the number of stars $N$.
(b1) Same as (a1), but the temperature of the heavy component $T_2$ is plotted.
(a2) and (b2) are the same as (a1) and (b1), but for model T2.
(a3) and (b3) are the same as (a1) and (b1), but for model T3.
}
\end{Fv}

\newpage

\begin{center}
\leavevmode
\epsfverbosetrue
\epsfxsize=14cm \epsfbox{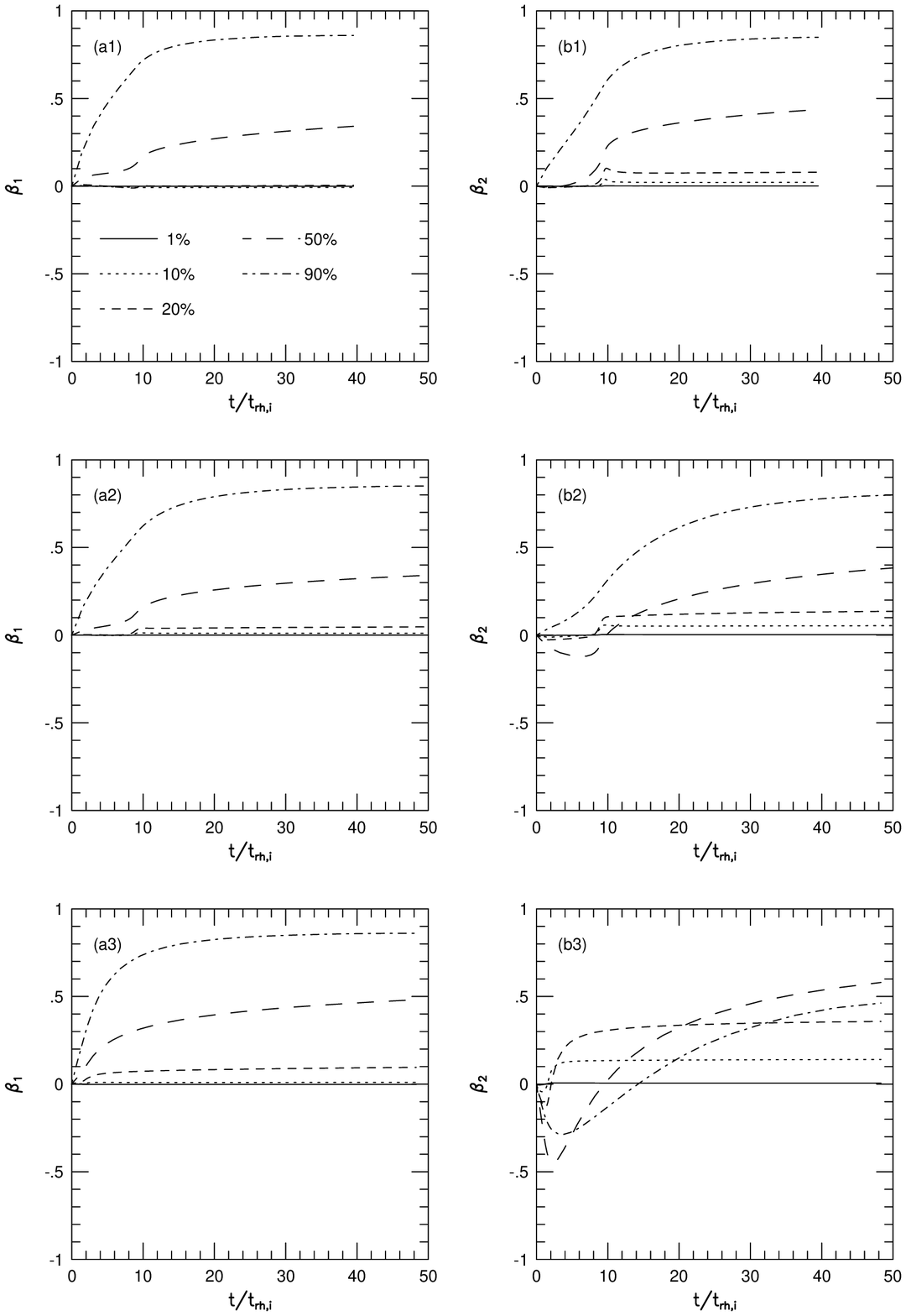}
\end{center}
\begin{Fv}{3}
{30pc}
{
(a1) Evolution of the anisotropy of the light component $\beta_1$
at the Lagrangian radii containing 1, 10, 20, 50, and 90\% of the total mass of the cluster, for model T1.
(b1) Same as (a1), but the anisotropy of the heavy component $\beta_2$ is plotted.
(a2) and (b2) are the same as (a1) and (b1), but for model T2.
(a3) and (b3) are the same as (a1) and (b1), but for model T3.
}
\end{Fv}

\newpage

\begin{center}
\leavevmode
\epsfverbosetrue
\epsfxsize=14cm \epsfbox{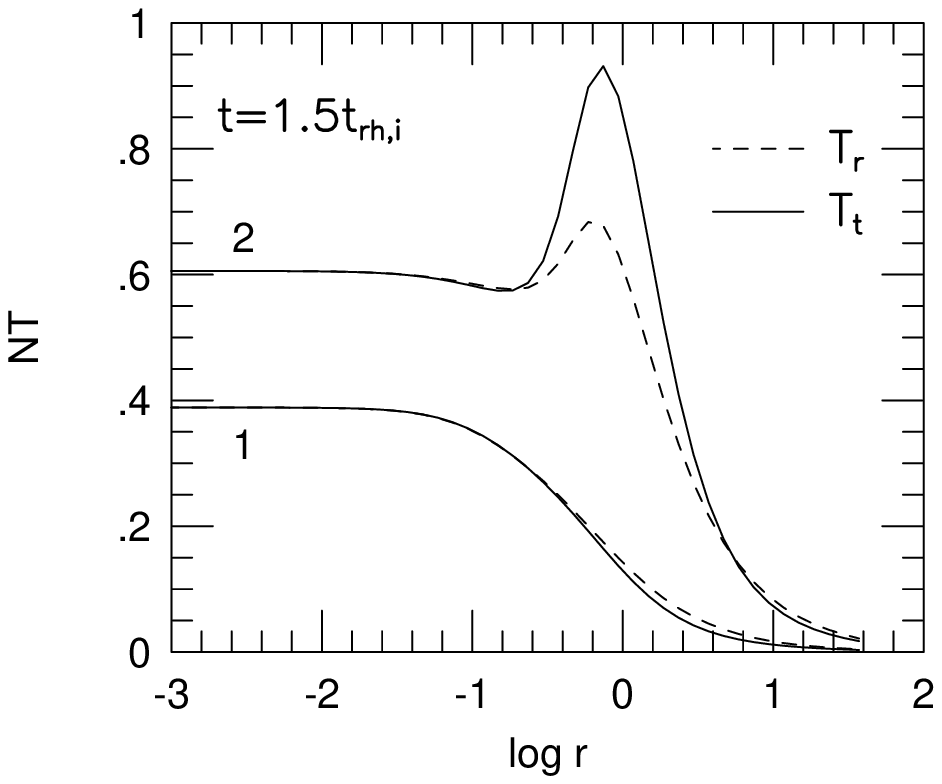}
\end{center}
\begin{Fv}{4}
{30pc}
{
Radial profiles of the radial temperature ($T_{\rm r}$) and
the tangential temperature ($T_{\rm t}$)
of each component at $t=1.5t_{\rm rh,i}$ in model T3.
Component numbers are indicated in the figure.
The ordinate is the temperature multiplied by the number of stars $N$.
}
\end{Fv}

\begin{center}
\leavevmode
\epsfverbosetrue
\epsfxsize=14cm \epsfbox{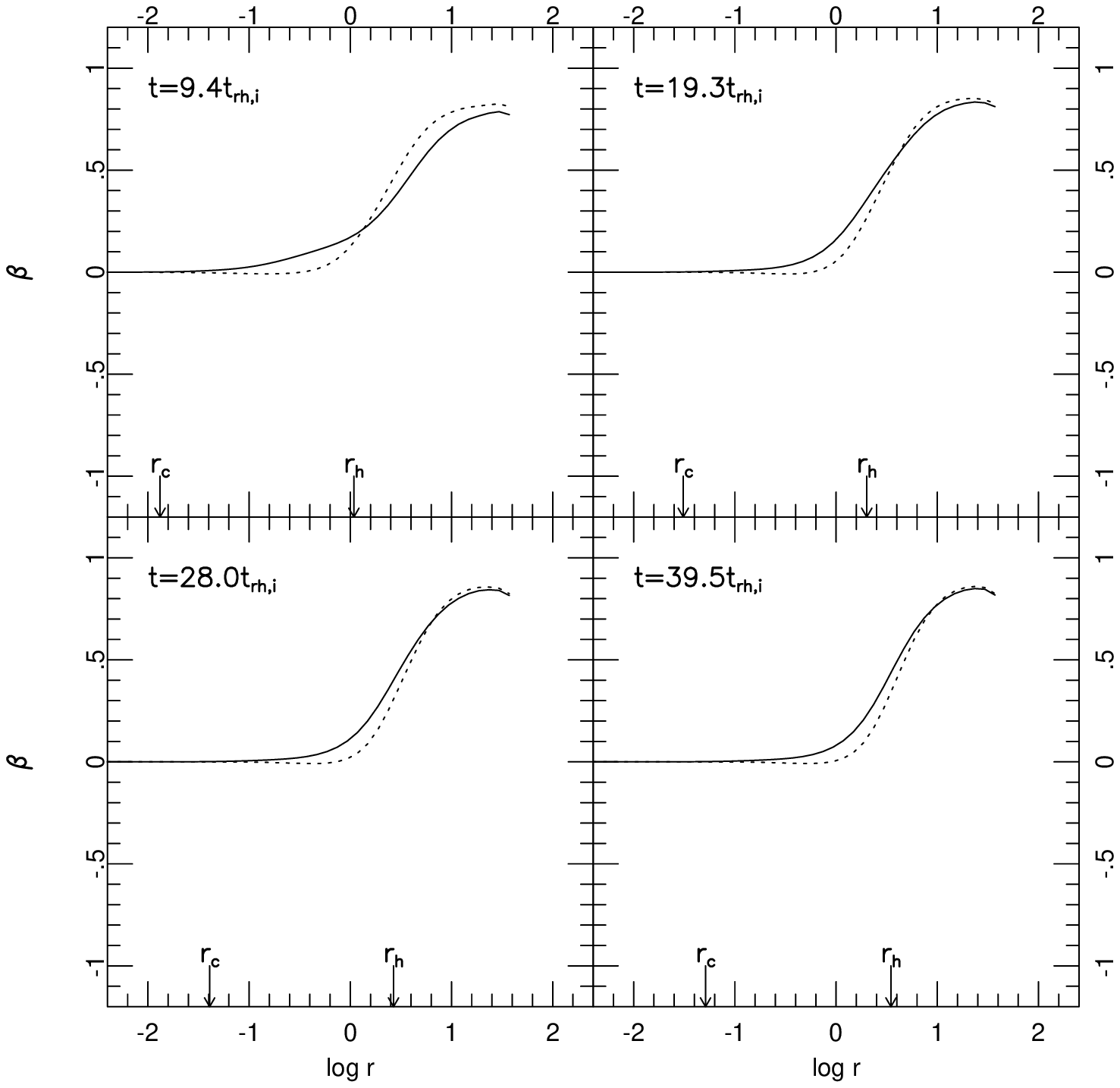}
\end{center}
\begin{Fv}{5}
{30pc}
{
Evolution of the radial profile of the anisotropy $\beta$ in model T1.
The profiles at four different epochs are shown. 
The solid and dotted lines represent the heavy and light components, respectively.
The core radius $r_{\rm c}$ and the half-mass radius $r_{\rm h}$ are also
indicated.
}
\end{Fv}

\begin{center}
\leavevmode
\epsfverbosetrue
\epsfxsize=14cm \epsfbox{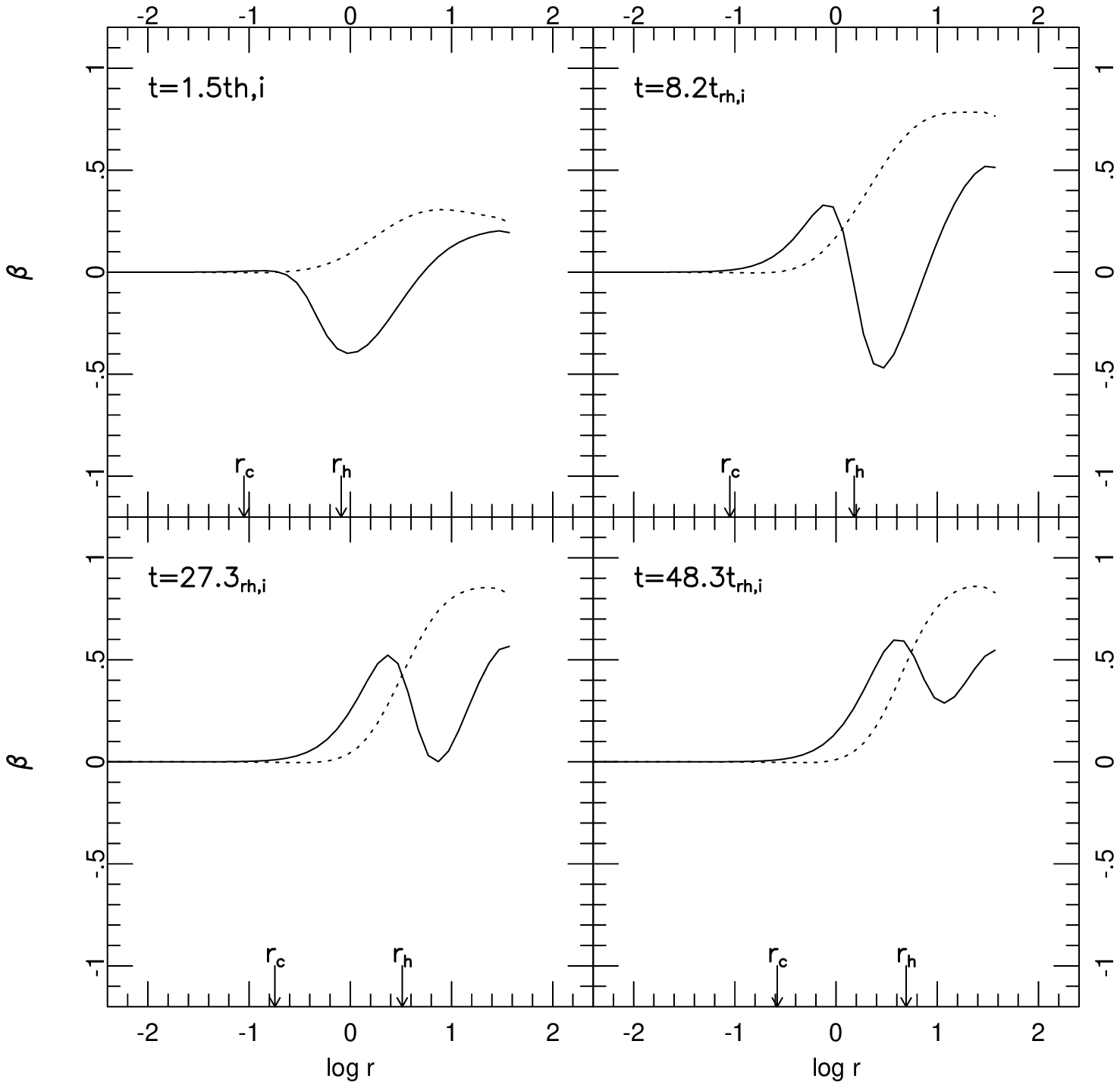}
\end{center}
\begin{Fv}{6}
{30pc}
{
Same as figure 5, but for model T3.
}
\end{Fv}

\begin{center}
\leavevmode
\epsfverbosetrue
\epsfxsize=14cm \epsfbox{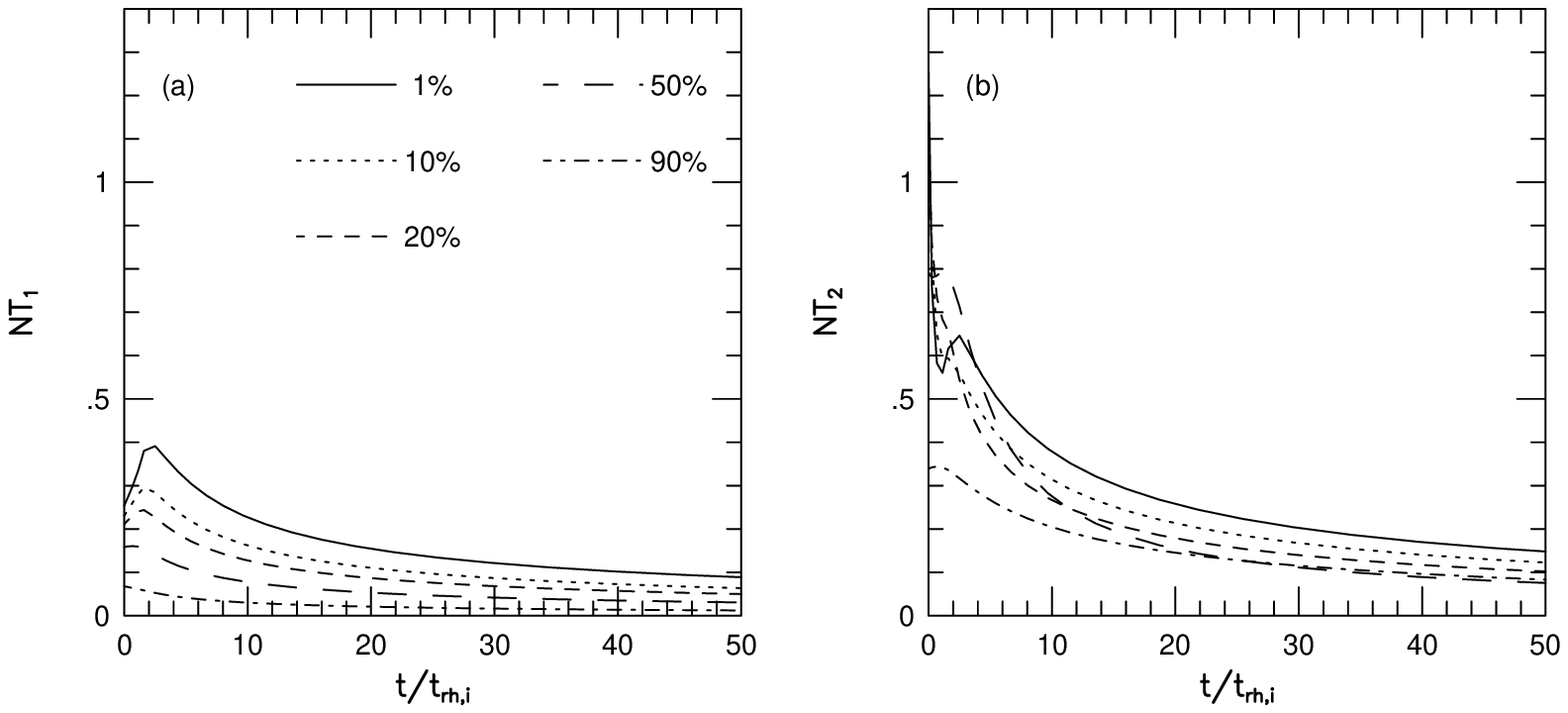}
\end{center}
\begin{Fv}{7}
{30pc}
{
(a) Evolution of the temperature of the light component $T_1$
at the Lagrangian radii containing 1, 10, 20, 50, and 90\% of the total mass of the cluster, for the isotropic model corresponding to model T3.
The ordinate is the temperature multiplied by the number of stars $N$.
(b) Same as (a), but the temperature of the heavy component $T_2$ is plotted.
}
\end{Fv}

\begin{center}
\leavevmode
\epsfverbosetrue
\epsfxsize=14cm \epsfbox{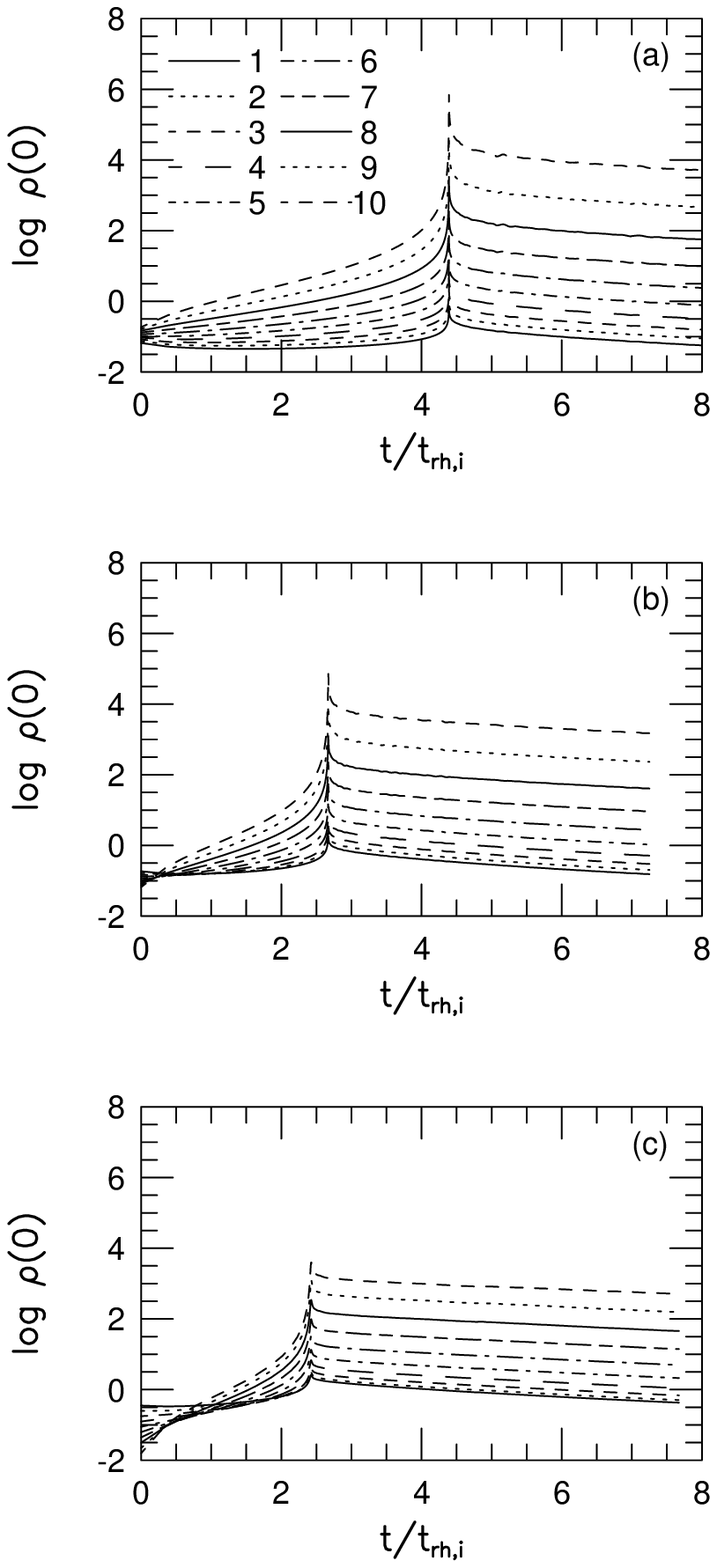}
\end{center}
\begin{Fv}{8}
{30pc}
{
Evolution of the central density of each component 
in models (a) C1, (b) C2, and (c) C3.
Component numbers are indicated in the figure. 
}
\end{Fv}

\begin{center}
\leavevmode
\epsfverbosetrue
\epsfxsize=14cm \epsfbox{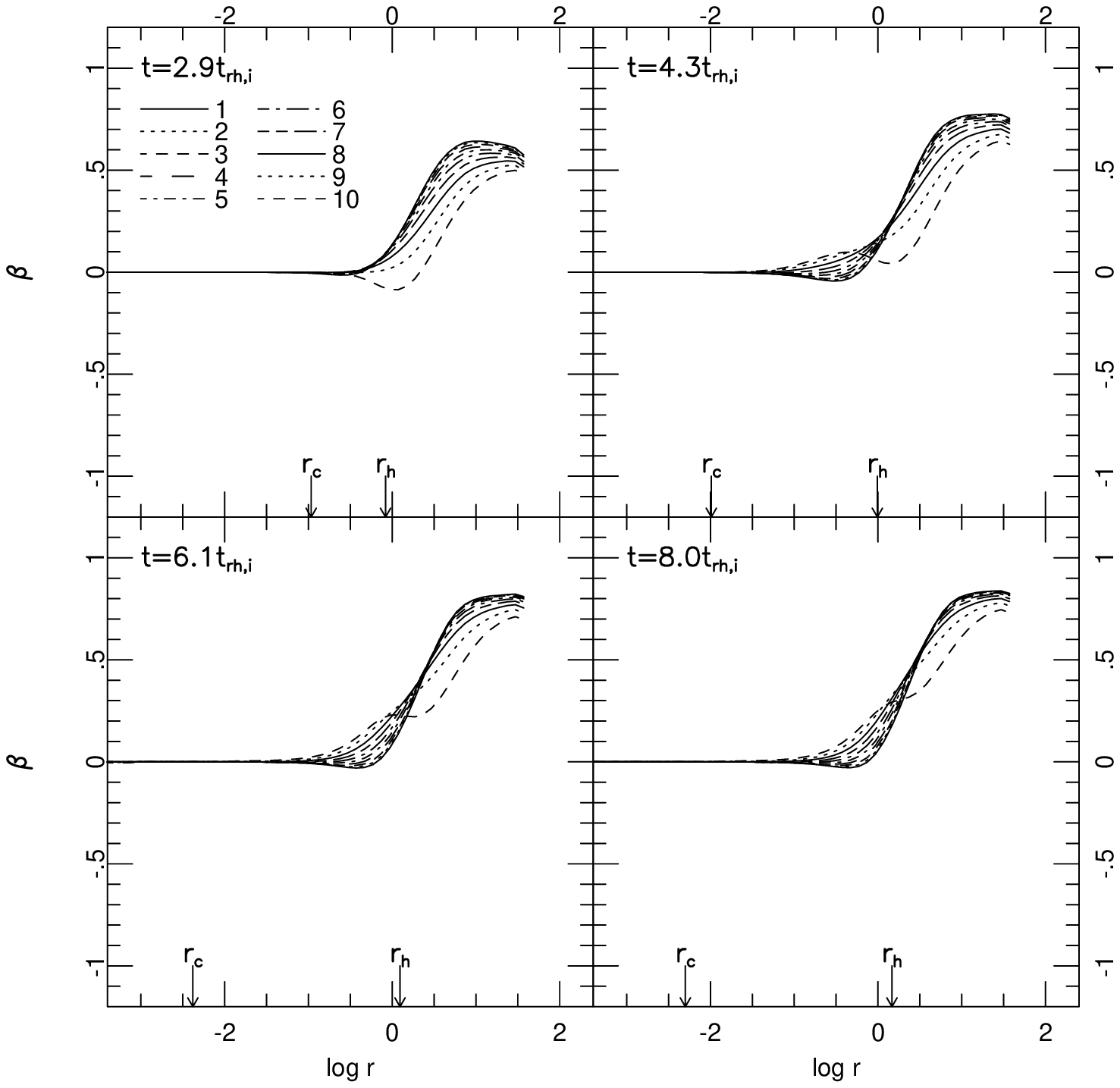}
\end{center}
\begin{Fv}{9}
{30pc}
{
Evolution of the anisotropy profile of each component in model C1.
Component numbers are indicated in the figure. 
The profiles at four different epochs are shown. 
The upper two panels are for the pre-collapse phase
and the lower two panels are for the post-collapse phase.
The core radius $r_{\rm c}$ and the half-mass radius $r_{\rm h}$ are also 
indicated.
}
\end{Fv}

\begin{center}
\leavevmode
\epsfverbosetrue
\epsfxsize=14cm \epsfbox{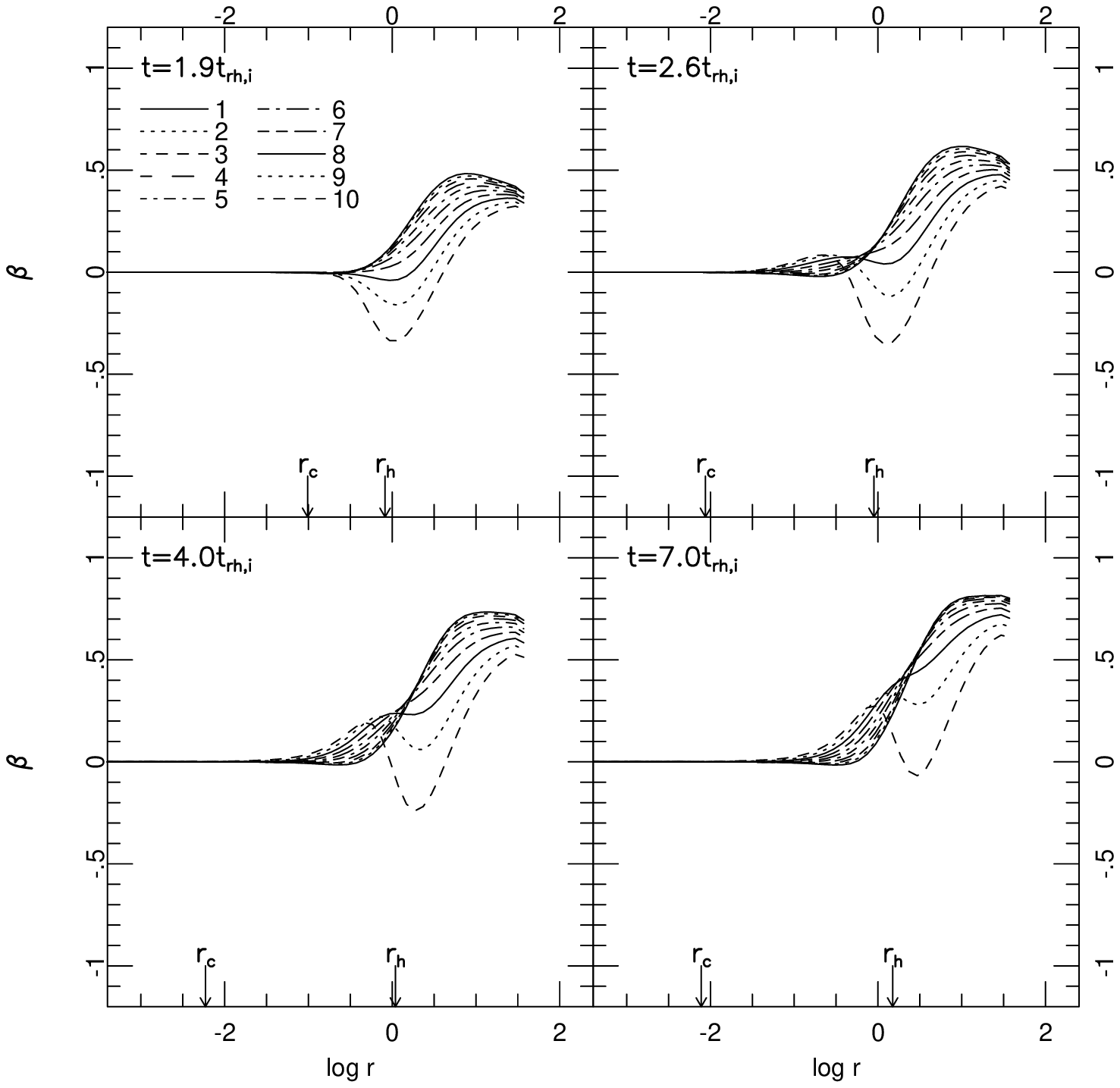}
\end{center}
\begin{Fv}{10}
{30pc}
{
Same as figure 9, but for model C2.
}
\end{Fv}

\begin{center}
\leavevmode
\epsfverbosetrue
\epsfxsize=14cm \epsfbox{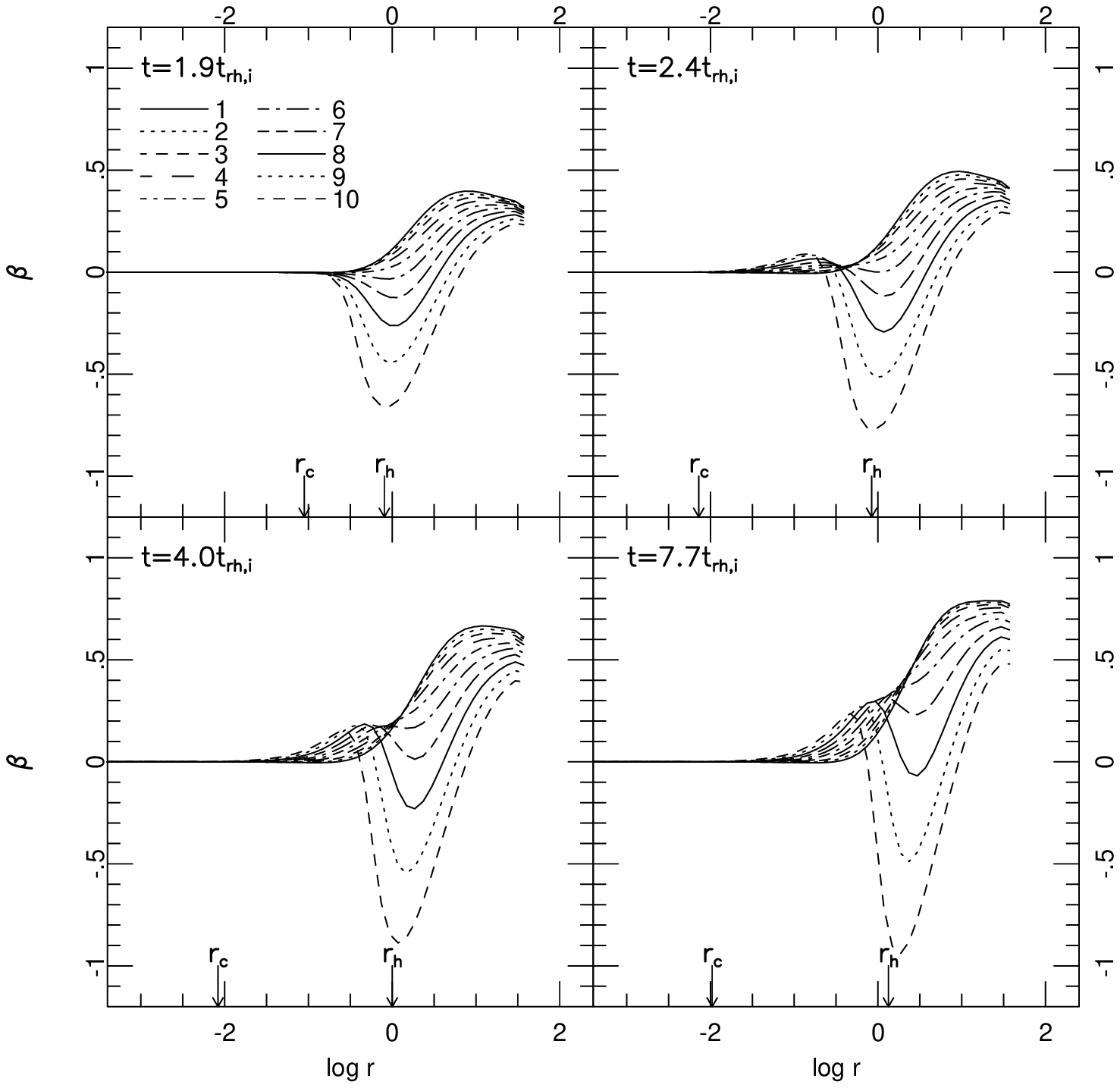}
\end{center}
\begin{Fv}{11}
{30pc}
{
Same as figure 9, but for model C3.
}
\end{Fv}

\begin{center}
\leavevmode
\epsfverbosetrue
\epsfxsize=14cm \epsfbox{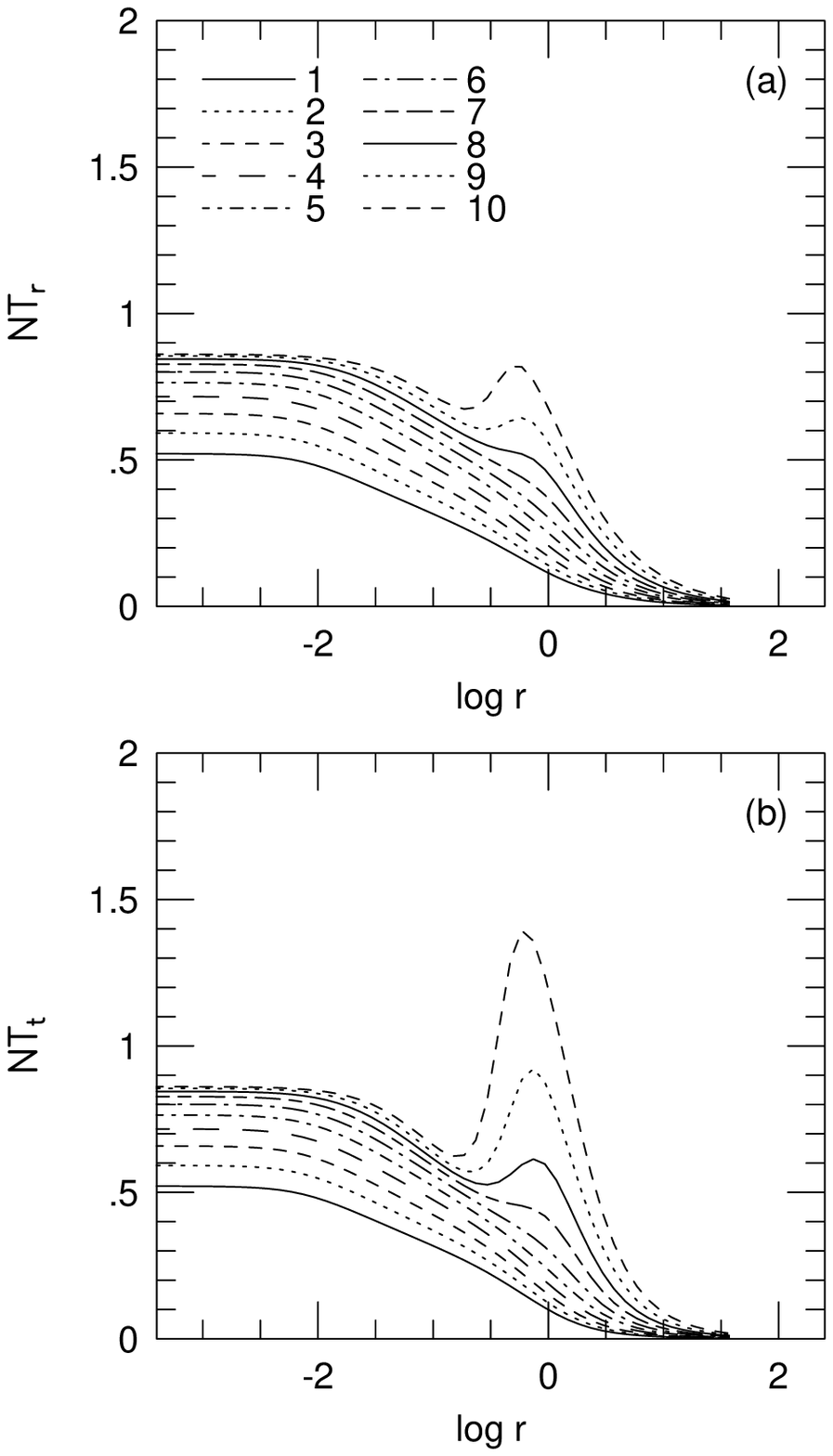}
\end{center}
\begin{Fv}{12}
{30pc}
{
Profiles of the (a) radial and (b) tangential temperatures of each component 
at $t=2.4t_{\rm rh,i}$ in model C3.
Component numbers are indicated in the figure. 
The ordinate is the temperature multiplied by the number of stars $N$.
}
\end{Fv}

\end{document}